\documentclass[10pt, conference, compsocconf]{IEEEtran}
\usepackage[pdftex]{graphicx}
\DeclareGraphicsExtensions{.png,.jpeg}
\usepackage[cmex10]{amsmath}
\usepackage[backend=bibtex,bibstyle=ieee,citestyle=numeric]{biblatex}
\usepackage{amsfonts}
\usepackage{array}
\usepackage{caption}
\bibliography{ipdps}

\begin{document}
	%
	\title{A High Performance Implementation of Spectral Clustering  on CPU-GPU Platforms}

	
	\author{\IEEEauthorblockN{Yu Jin}
		\IEEEauthorblockN{Institute for Advanced Computer Studies \\ Department of Electrical and Computer Engineering\\ University of Maryland, College Park, USA\\
			Email: yuj@umd.edu\\}
		\and \IEEEauthorblockN{Joseph F. JaJa}
		\IEEEauthorblockN{Institute for Advanced Computer Studies \\ Department of Electrical and Computer Engineering\\ University of Maryland, College Park, USA\\
			Email: joseph@umiacs.umd.edu \\}
	}

	\maketitle

	\begin{abstract}
		Spectral clustering is one of the most popular graph clustering algorithms, which achieves the best performance for many scientific and engineering applications. However, existing implementations in commonly used software platforms such as Matlab and Python do not scale well for many of the emerging Big Data applications. In this paper, we present a fast implementation of the spectral clustering algorithm on a CPU-GPU heterogeneous platform. Our implementation takes advantage of the computational power of the multi-core CPU and the massive multithreading and SIMD capabilities of GPUs. Given the input as data points in high dimensional space, we propose a parallel scheme to build a sparse similarity graph represented in a standard sparse representation format. Then we compute the smallest $k$ eigenvectors of the Laplacian matrix by utilizing the reverse communication interfaces of ARPACK software and cuSPARSE library, where $k$ is typically very large. Moreover, we implement a very fast parallelized $k$-means algorithm on GPUs. Our implementation is shown to be significantly faster compared to the best known Matlab and Python implementations for each step. In addition, our algorithm scales to problems with a very large number of clusters.
	\end{abstract}
	

	%
	\IEEEpeerreviewmaketitle

	\section{Introduction}
	Spectral clustering algorithm has recently gained popularity in handling many graph clustering tasks such as those reported in \cite{van2013community, jin2015, craddock2012whole}. Compared to traditional clustering algorithms, such as k-means clustering and hierarchical clustering, spectral clustering has a very well formulated mathematical framework and is able to discover non-convex regions which may not be detected by other clustering algorithms. Moreover, spectral clustering can be conveniently implemented by linear algebra operations using popular scientific software environments such as Matlab and Python. Most of the available software implementations are built upon CPU-optimized Basic Linear Algebra Subprograms (BLAS), usually accelerated using multi-thread programming. However, such implementations scale poorly as the problem size or the number of clusters grow very large. Recent results show that GPU accelerated BLAS significantly outperforms multi-threaded BLAS libraries such as  the Intel MKL package, LAPACK and Goto BLAS \cite{eddelbuettel2010benchmarking, cullinan2013computing}. Moreover, hybrid computing environments, which collaboratively combine the computational advantages of GPUs and CPUs, further boost the overall performance and are able to achieve very high performance on problems whose sizes grow up to the capacity of CPU memory \cite{lee2014boosting, agulleiro2012hybrid, wu2014optimized, wuachieving, lihybrid, akimova2012parallel}. 
	In this paper, we present a hybrid implementation of the spectral clustering algorithm which significantly outperforms the known implementations, most of which are purely based on multi-core CPUs.  
	
	\par There have been reported efforts on parallelizing the spectral clustering algorithm. Zheng et al.  \cite{zheng2008parallelization} presented both CUDA and OpenMP implementations of spectral clustering. However, the implementation was targeted for a much smaller data size than the work in this paper, and moreover, their implementation achieve a relatively limited speedup. Matam et al. \cite{matam2011gpu} implemented a special case of spectral clustering, namely the spectral bisection algorithm, which was shown to achieve high speed-ups compared to Matlab and Intel MKL implementations. Chen et al.  \cite{Chen11, song2008parallel} implemented the spectral clustering algorithm on a distributed environment using Message Passing Interface (MPI), which is targeted for problems whose sizes that could not fit in the memory of a single machine. Tsironis and Sozio \cite{tsironis2013accurate} proposed an implementation of spectral clustering based on MapReduce. Both implementations were targeted for clusters, and involve frequent data communications which will clearly constrain the overall performance.

	\par In this paper, we present a hybrid implementation of spectral clustering on a CPU-GPU heterogeneous platform which significantly outperforms all the best implementations we are aware of, which are based on existing parallel platforms. We highlight the main contributions of our paper as follows:
	
	\begin{itemize}
		\item Our algorithm is the first work to comprehensively explore the hybrid implementation of spectral clustering algorithm on CPU-GPU platforms.
		\item Our implementation makes use of sparse representation of the corresponding graphs and can handle extremely large input sizes and generate a very large number of clusters. 
		\item The hybrid implementation is highly efficient and is shown to make a very good use of available resources.
		\item Our experimental results show superior performance relative to the common scientific software implementations on multicore CPUs.
	\end{itemize}
	
	\par The rest of the paper is organized as follows. Section II gives an overview of the spectral clustering algorithm, while describing the important steps in some detail. Section III describes the operating environment and the necessary software dependencies. Section IV provides a description of our parallel implementation, while Section V evaluates the performance of our algorithm with a comparison with Matlab and Python implementations on both synthetic and real-world datasets. The codes are available on https://github.com/yuj-umd/fastsc.
	
	\section{Overview of Spectral Clustering Algorithm}
	Spectral clustering was first introduced in 1973 to study the graph partition problem \cite{donath1973lower}. Later, the algorithm was extended in \cite{shi2000normalized, ng2002spectral}, and generalized to a wide range of applications, such as computational biology \cite{pentney2005spectral, higham2007spectral}, medical image analysis \cite{jin2015, craddock2012whole}, social networks \cite{white2005spectral, mishra2007clustering} and information retrieval \cite{chifu2015word, mcfee2014analyzing}. A  standard procedure of the  spectral clustering algorithm to compute $k$ clusters is described next \cite{von2007tutorial},
	
	\begin{itemize}
		\item Step 1: Given a set of data  points $x_1, x_2, ... , x_n \in \mathbb{R}^d$ and some similarity measure $s(x_i,x_j)$, construct a sparse similarity matrix $W$ that captures the significant similarities between the pairs of points. 
		\item Step 2: Compute the normalized graph Laplacian matrix as $L_n = D^{-1}L$ where $L$ is the unnormalized graph Laplacian matrix defined as $L = D - W$ and $D$ is the diagonal matrix with each element $D_{i,i} = \sum_{j=1 }^{n} W_{i,j}$.
		\item Step 3: Compute the $k$ eigenvectors of the normalized graph Laplacian matrix $L_n$ corresponding to the smallest $k$ nonzero eigenvalues. 
		\item Step 4: Apply the  $k$-means clustering algorithm on the rows of the matrix whose columns are the $k$ eigenvectors to obtain the final clusters.
	\end{itemize} 
	
	\par Given the similarity graph defined by the similarity matrix $W$, the basic idea behind spectral clustering is to partition the graph into $k$ partitions such that some measure of the cut between the partitions is minimized. The traditional graph cut is defined as follows:
	
	\begin{equation}
	\label{cut}
	\text{Cut}(A_1, A_2,..., A_k) = \frac{1}{2} \sum_{i = 1}^{k} W(A_i, \bar{A_i});
	\end{equation}
	
	\begin{equation}
	W(A, \bar{A}) := \sum_{i \in A, j \in \bar{A}} w_{ij}
	\end{equation}
	
	To ensure that the each partition represents a meaningful cluster of reasonable size, two alternative cut measures are often used, namely \textbf{RatioCut} and normalized cut \textbf{Ncut}. Note that we use below $\lvert A_i \lvert$ as the number of nodes in $A$ and $vol(A)$ as the sum of the degrees of all the nodes in $A$.
	
	\begin{equation}
	\text{RatioCut}(A_1, A_2, A_k) = \frac{1}{2} \sum_{i = 1}^{k} \frac{W(A_i, \bar{A_i})}{\lvert A_i \lvert};
	\end{equation}
	\begin{equation}
	\text{Ncut}(A_1, A_2, A_k) = \frac{1}{2} \sum_{i = 1}^{k} \frac{W(A_i, \bar{A_i})}{vol(A_i)};
	\end{equation}
	In our implementation, we focus on the problem of minimizing the \textbf{Ncut} which has an equivalent algebraic formulation as defined next.
	\begin{equation}
	\begin{aligned}
	& \underset{H}{\text{min}} \hspace{1.0mm} \mathrm{trace}(H'LH) \hspace{1.0mm} \text{subject to}\hspace{1.0mm} H'DH = I
	\end{aligned}
	\end{equation}
	
	That is, we need to determine a matrix  $H \in \mathbb{R}^{n \times k}$ whose columns are indicator vectors, which minimizes the objective function introduced above. 
	
	
	\par Since this problem is NP-hard, we relax the discrete constraints on $H$ are removed, thereby allowing $H$ to be any matrix in  $\mathbb{R}^{n \times k}$. Note that there is no theoretical guarantee on the quality of the solution of the relaxed problem compared to the exact solution of the discrete version. It turns out that the relaxed problem is a well-known trace minimization problem, which can be exactly solved by taking $H$ as the eigenvectors with the smallest $k$ eigenvalues of the matrix $L_n = D^{-1}L$ or equivalently the $k$ generalized eigenvectors corresponding to the smallest $k$ eigenvalues of $Lx = \lambda Dx$.  The k-means clustering is then applied on the rows of $H$ to obtain the desired clustering. 
	
	\par The algorithm described above begins with a set of $d$-dimensional data points and builds the similarity graph explicitly from the pair-wise similarity metric. The similarity graph is usually stored in a sparse matrix representation, which often reduces the memory requirement and computational cost to linear instead of quadratic.  For the general graph clustering whose input is specified as a graph, our spectral clustering algorithm starts directly in Step 2. Otherwise, we build our sparse graph representation from the given set of data points.
	

	\section{Environment Setup}
	\subsection{The Heterogeneous System}
	The CPU-GPU heterogeneous system 
	used in our implementation is specified in Table \ref{Specifics}.
	
	\begin{table}[t!]
		\caption{CPU and GPU specifics}
		\label{Specifics}
		\centering
		\begin{tabular}{|c|c|}
			\hline
			CPU Model & Intel Xeon E5-2690 \\
			\hline 
			CPU Cores & 8 \\
			\hline 
			DRAM Size & 128GB \\
			\hline 
			GPU Model & Tesla K20c\\
			\hline 
			Device Memory Size & 5GB GDDR5 \\
			\hline
			SMs and SPs & 13 and 192\\
			\hline 
			Compute Capability & 3.5\\
			\hline
			CUDA SDK & 7.5 \\
			\hline 
			PCIe Bus & PCIe x16 Gen2 \\
			\hline
		\end{tabular}
	\end{table}
	
	The CPU and the GPU communicate through the PCIe bus whose theoretical peak bandwidth is 8 GB/s. The cost of data communication can be quite significant for large-scale problems. To achieve the best overall performance, our implementation leverages the GPU to compute the most computationally expensive part while minimizing the data transfer between the host and the device.

	\subsection{CUDA Platform}
\begin{figure}[!t]
		\centering
		\includegraphics[width = .8\hsize]{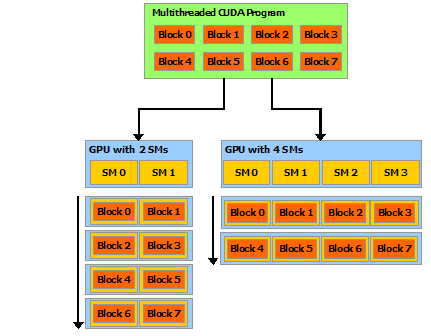}
		\caption{CUDA Program Model}
		\label{cuda}
	\end{figure}

	CUDA is a general-purpose multithreaded programming model that leverages the large number of GPU cores to solve complex data parallel problems. The CUDA programming model assumes a heterogeneous system with a host CPU and several GPUs as co-processors. Each GPU has an array of Streaming Multiprocessors (\textbf{SM}), each of which has a number of Streaming Processors (\textbf{SP}) that execute instructions concurrently. The parallel computation on GPU is invoked by calling customized kernel functions using thousands of threads. The kernel function is executed by blocks of threads independently. Each block of threads can be scheduled on any Streaming Multiprocessors (SP) as shown in Figure \ref{cuda}. The kernel function takes as  parameters  the number of blocks and the number of threads within a block.
	
	\par In addition, NVIDIA provides efficient BLAS libraries for both sparse\footnote{http://docs.nvidia.com/cuda/cusparse/} and dense\footnote{http://docs.nvidia.com/cuda/cublas/} matrix computations. Our implementation relies on the Thrust library, which resembles the C++ Standard Template Library (STL) that provides efficient operations such as sort, transform, which greatly improves productivity.

	\subsection{ARPACK Software}
	ARPACK is a software package designed to solve large-scale eigenvalue problems \cite{lehoucq1997arpack}. ARPACK is reliable and achieves high accuracy, and is widely used in modern scientific software environments. It contains highly optimized Fortran subroutines that are able to solve symmetric, non-symmetric and generalized eigenproblems. ARPACK is based on the Implicitly Restarted Arnoldi Method (IRAM) with non-trivial numerical optimization techniques \cite{lehoucq1996deflation, sorensen1992implicit}.  In our implementation, we adopt ARPACK++ \footnote{http://reuter.mit.edu/software/arpackpatch/} that provides C++ interfaces to the original ARPACK Fortran packages and utilizes efficient matrix solver libraries such as LAPACK, SuperLU. The eigenvalue problem is efficiently solved by collaboratively combining the interfaces of ARPACK++ and cuSPARSE library.
	\subsection{OpenBLAS}
	OpenBLAS\footnote{http://www.openblas.net/} is an open-source CPU-based BLAS library utilized by ARPACK++. It supports multi-threaded acceleration through pthread programming or OpenMP by specifying corresponding environment variables. OpenBLAS is a highly optimized BLAS library developed based on GotoBLAS2, which has been shown to surpass other CPU-based BLAS libraries \cite{eddelbuettel2010benchmarking}.
	
	\section{Implementation}
	\subsection{Data Preprocessing}
	Given the $d$-dimensional data points, the preprocessing step constructs the similarity matrix from the data points. The clustering problem is reformulated as a graph clustering where the graph is represented by the similarity matrix.
	
	\par As mentioned before, the similarity matrix is usually constructed to be sparse, which reduces the memory requirement and enables high computational efficiency. The sparsity patterns of the similarity matrices are highly dependent on the specific application. The following are several common ways to construct a sparse similarity matrix \cite{von2007tutorial}.

	\begin{itemize}
		\item $\lambda$-threshold graph: The similarity graph is constructed where data points are connected if their similarity measure is above the threshold $\lambda$.
		\item $\varepsilon$-distance graph: The similarity matrix is construct by only connecting data points that are within a spatial distance $\varepsilon$. 
		\item $k$-nearest-neighbor graph: The similarity graph is constructed where two data points $x_i$ and $x_j$ are connected only if either $x_i$ is among the $k$ most similar data points of $x_j$, or $x_j$ is among the  $k$ most similar data points of $x_i$. Note that the parameter $k$ is unrelated to the number $k$ of clusters used in the next section.
	\end{itemize}
	
	\par The notion of the similarity measure between data points also varies depending on the application. Typical measures are the following.
	
	\begin{itemize}
		\item Cosine Similarity Measure
		\begin{equation}
		\text{CosineDist}(x_i, x_j) = \frac{\langle x_i, x_j\rangle}{\Vert x_i \Vert_2\Vert x_j \Vert_2}
		\end{equation}
		\item Cross Correlation 
		\begin{equation}
		\text{CrossCorr}(x_i, x_j) = \frac{\langle x_i - \bar{x_i}, x_j - \bar{x_j}\rangle}{\Vert x_i- \bar{x_i} \Vert_2\Vert x_j - \bar{x_j} \Vert_2}
		\end{equation}
		\item Exponential decay function
		\begin{equation}
		\text{ExpDecay}(x_i, x_j) = e^{\frac{\Vert x_i- x_j \Vert_2}{2\sigma^2}}
		\end{equation}
	\end{itemize}
	
	\par Although the sparse patterns and similarity measures are different depending on the application, the general construction of the similarity matrix can be accelerated under the CUDA programming model regardless of the preprocessing used. Here we provide a parallel implementation for a specific sparsity pattern and similarity measure.
	
	\par We consider the input data as a matrix $X \in \mathbb{R}^{n\times d}$ where $n$ is the number of data points and $d$ is the dimension of each data point. The goal is to construct a sparse matrix representation of the similarity graph using the $\varepsilon$-distance graph structure and cross correlation as the similarity measure. We assume the neighborhood information is given by a  list $E \in \mathbb{R}^{nnz\times 2}$, which contains all pairs of indices of data points that are  within $\varepsilon$-distance. The number $nnz$ of such pairs is the number of edges in the graph. The procedure for constructing the sparse similarity matrix represented in Coordinate Format (COO) format is described in Algorithm 1.
	
	\begin{table}[t!]
		\centering
		\begin{tabular}{m{8cm}}
			\hline \hline
			\textbf{Algorithm 1} Construction of Sparse Similarity Matrix \\
			\hline
			1. Transfer the input data $X$ and edge lists $E$ from CPU to GPU.\\
			2. Initialize $n$-length vectors \textbf{$X_{average}$} and \textbf{$X_{norm}$} on GPU.\\
			3. Initialize $nnz$-length vector $val$ on GPU.\\
			4. Execute kernel function \texttt{compute\_average} where each thread $i$  computes $X_{average}(i) = \frac{1}{d}\sum_{j = 1}^{d}X_{ij}$ \\
			5. Execute kernel function \texttt{update\_data} where each thread $i$ updates one row of data $X_{ij} = X_{ij} - X_{average}(i)$ and compute $X_{norm}(i) = \sqrt{\sum_{j = 1}^{d}X_{ij}^2}$\\
			6. Execute kernel function \texttt{compute\_similarity} where each thread $i$ computes the similarity between the $i^{th}$ pair of data points in $E$. \\
			7. The edge list and the vector $val$ form the sparse graph represented in the Coordinate Format (COO) format.\\
			\hline \hline	
		\end{tabular}
	\end{table}
	
	The above procedure is highly data parallel and  easy to implement under the CUDA programming model. In general, there are two sparse matrix representations that we use in our work. 
	\begin{itemize}
		\item Coordinate Format (\textbf{COO}): this format is the simplest sparse matrix representation. Essentially, COO uses 
		tuples $(i, j, w_{ij})$ to represent all the non-zero entries. This can be done through three separate $nnz$-length arrays that respectively store the row indices, column indices, and the corresponding non-zero matrix values.
		\item Compressed Sparse Row Format (\textbf{CSR}): this consists of three arrays, one containing the non-zero values, 
		the second containing the column indices of the corresponding non-zero values, and the third contains the prefix sums of the 
		number of nonzero entries of the rows. 
	\end{itemize}
	
	Other sparse formats such as Compressed Sparse Column Format (\textbf{CSC}), Block Compressed Sparse Row Format (\textbf{BSR}) are also supported in our implementation. 
	
	\subsection{Parallel Eigensolvers}
	Given the similarity graph $W$ represented in some sparse format and the desired number of clusters $k$, this step computes the $k$ eigenvectors corresponding to the smallest $k$ eigenvalues of normalized Laplacian $L_n = I - D^{-1}W$
	where $W$ is the sparse matrix and $D$ is the diagonal matrix with each element $D_{i,i} = \sum_{j=1 }^{n} W_{i,j}$. We assume that $D_{i,i}$ are all positive, otherwise the isolated nodes can be removed from the graph. The eigenvectors corresponding to the smallest $k$ eigenvalues of the normalized Laplacian are exactly the eigenvectors corresponding to the largest $k$ eigenvalues of $D^{-1}W$. Since computing the largest eigenvalues results in better numerical stability and convergent behavior, we focus our attention on computing the eigenvectors corresponding to the largest $k$ eigenvalues of $D^{-1}W$. 
	
	\par The sparse matrix multiplication $D^{-1}W$ can easily be computed as follows:
	\begin{equation}
	\left[
	\begin{tabular}{m{0.3cm}m{0.3cm}m{0.3cm}m{0.3cm}}
	$d_{11}^{-1}$ & & & \\
	& $d_{22}^{-1}$ & & \\
	& & ... & \\
	& & & $d_{nn}^{-1}$\\
	\end{tabular}
	\right] \times 
	\left[
	\begin{tabular}{m{0.5cm}}
	$W_{1j}$ \\
	$W_{2j}$ \\
	... \\
	$W_{nj}$
	\end{tabular}
	\right] = 
	\left[
	\begin{tabular}{m{1.0cm}}
	$d_{11}^{-1}W_{1j}$ \\
	$d_{22}^{-1}W_{2j}$ \\
	... \\
	$d_{nn}^{-1}W_{nj}$
	\end{tabular}
	\right]
	\end{equation}
	
	\par The corresponding computation is data parallel and has complexity O($nnz$). We assume that the sparse similarity matrix initially resides in the device memory, represented in \textbf{COO} format. The parallel computation is described in Algorithm 2. Note that the $D^{-1}W$ will be transformed to the \textbf{CSR} format to perform the sparse matrix-vector multiplication at the next step.
	
	\begin{table}[t!]
		\centering
		\begin{tabular}{m{8cm}}
			\hline \hline
			\textbf{Algorithm 2} Parallel Computation of $D^{-1}W$ \\
			1. Initialize a n-length vector $x$ with 1.0 for all elements. \\
			2. Compute the vector $y = Wx$ where each element $y_i = d_{ii}$ by calling \texttt{cusparseDcsrmv} in cuSPARSE library\\
			3. Execute the kernel function \texttt{ScaleElements} where each thread i processes one item in COO format $<\text{r}, \text{c}, \text{val}>$ and scales the element value by the inverse of $y_i$.\\ 
			4. Compress the row indices through the cuSPARSE interface \texttt{cusparseXcoo2csr}. \\
			5. The compressed row indices, the column indices and the updated element value form the CSR representation of $D^{-1}W$ \\
			\hline
			\hline \hline	
		\end{tabular}
	\end{table}
	
	\par An important feature of the ARPACK software is the \textbf{reverse communication interfaces}, which facilitate the process of solving large-scale eigenvalue problems. The reverse communication interfaces are CPU-based interfaces that encapsulate implicitly restarted Arnoldi/Lanczos method, which is an iterative method to obtain the required eigenvalues and corresponding eigenvectors. For each iteration, the interface provides a $n$-length vector used as input and the output of sparse matrix-vector multiplication is provided back to the interface. ARPACK interfaces combine the optimized Fortran routines and CPU-based BLAS library OpenBLAS, which is one of the most efficient CPU-based BLAS library. ARPACK provides the flexibility in choosing any matrix representation format and the function to obtain the results of matrix-vector multiplication. In our implementation, the matrix-vector multiplication is performed on the GPU. For each iteration, the input vector is transferred from the CPU to the GPU and the output vector is transfered back to the interface. The detailed implementation is shown in Algorithm 3.

	\begin{table}[t!]
		\centering
		\begin{tabular}{m{8cm}}
			\hline \hline
			\textbf{Algorithm 3} Parallel Eigensolver \\
			\hline
			1. Initialize the object \texttt{Prob} with parameters. \\
			2. While \texttt{!Prob.converge()}\\
			\hspace{0.2cm} \texttt{Prob.TakeStep().} \\
			\hspace{0.2cm} Transfer the data located at \texttt{Prob.GetVector()} from host to device. \\
			\hspace{0.2cm} Call \texttt{cusparseDcsrmv} to perform matrix-vector multiplication on device.\\
			\hspace{0.2cm} Transfer the result from device to host and put it at the location addressed by \texttt{Prob.PutVector()}.\\
			3. Compute the eigenvectors by \texttt{Prob.FindEigenvectors()}.  \\	
			\hline \hline	
		\end{tabular}
	\end{table}

	\par The object \texttt{Prob}  is initialized as the eigenvalue problem for the symmetric real matrix with the $k$ largest-magnitude eigenvalues. \texttt{TakeStep()} is an interface that performs the necessary matrix operations based on the multi-threaded OpenBLAS library. For each iteration, the multiplication of sparse matrix and dense vector is computed on the GPU where 
	1) the sparse matrix is $D^{-1}W$ reside on GPU; 2) the input vector, whose location is indicated by \texttt{Prob.GetVector()}, is transferred from CPU to GPU; 3) the result is transfered back from GPU to CPU to the position \texttt{Prob.PutVector()}. After the object \texttt{Prob} reaches convergence, the eigenvectors are computed by \texttt{Prob.FindEigenvectors()}. 
	
	\par The complexity of Algorithm 3. largely depends on the interfaces \texttt{TakeStep()} and \texttt{FindEigenvectors()}. Both routines depend on the number  $m$ of Arnoldi/Lanczos vectors, which is usually set as $m = \text{max}(n, 2k)$. \texttt{TakeStep()} involves the eigenvalue decomposition and iteratively QR factorization of $m \times m$ matrix, as well as a few dense matrix-vector multiplication. Therefore the complexity for \texttt{TakeStep()} is at least $(O(m^3) + O(nm) \times O(m-k) )$. Moreover, the general complexity for sparse matrix-vector multiplication is $O(nnz\cdot m)$. The number of iteration $\#$ depends on the initial vector and properties of the matrix. The complexity \texttt{FindEigenvectors()} is $O(nmk)$. Hence the overall complexity is,
	\begin{equation}
	(O(m^3) + O(nm^2) + O(nnz \cdot m)) \times \# + O(nmk)
	\end{equation}

	\par As far as we know, the procedures described in Algorithm 3 are currently the most efficient and convenient way to solve general eigenvalue problems for large-scale matrices. We  leverage the existing software ARPACK on CPU to perform the 
	complex eigensolver procedures and the GPU to perform the expensive matrix computations. Results in Section V. will show that the data communication overhead is negligible compared to the overall computational cost and the overall implementation is very efficient compared to other software that relies on CPU-based sparse matrix-vector multiplication. 
	
	\subsection{Parallel k-means clustering}
	The k-means clustering algorithm is an iterative algorithm to partition the input data points into k clusters whose objective function is to minimize the sum of squared distances between each point and its representative. In spectral clustering, the k-means algorithm  is used to cluster the rows of the matrix consisting of the eigenvectors. Each such row can in fact be viewed as 
	a reduced dimension representation of the original data point. There are several GPU-based implementations of the k-means clustering such as \cite{zechner2009accelerating, wu2011efficient}. However, none of these implementations seem 
	to be efficient for large-scale problems, especially when k is very large. Our implementation is a revised version 
	from an open-source project \footnote{https://github.com/bryancatanzaro/kmeans} which efficiently utilizes the Thrust and CUBLAS libraries and achieve significant speedups.
	
	\par We assume that the low-dimensional representation $ V \in \mathbb{R}^{n \times k}$ initially resides in  the CPU memory where $n$ is the number of data points and $k$ is the desired number of clusters. The implementation is described in Algorithm 4.
	
	\begin{table}[t!]
		\centering
		\begin{tabular}{m{8cm}}
			\hline \hline
			\textbf{Algorithm 4} Parallel K-means Algorithm \\
			1. Transfer the data $V \in \mathbb{R}^{n \times d}$ from the CPU to the GPU. \\
			2. Randomly select $k$ points as the centroids of the k clusters stored in $C \in \mathbb{R}^{k \times d}$\\
			3. While (the centroids change) do \\
			\hspace{0.2cm} Compute the pairwise distances $S \in \mathbb{R}^{n \times k} $ between data points and the centroids.\\
			\hspace{0.2cm} Update the new label of each data point. \\
			\hspace{0.2cm} Compute the new centroids of the clusters.\\
			4. Transfer the labeling result from GPU to CPU.\\
			\hline 
			\hline \hline	
		\end{tabular}
	\end{table}
	
	\par Step 2 is the most common way to initialize the centroids. However, we use a more effective  initialization strategy, referred to as the  k-means++ initialization, which has been shown to converge faster and achieve better results than the traditional k-means algorithm \cite{arthur2007k}. This initialization is simple to implement in parallel using basic routines in CUDA Thrust library, as described in Algorithm 5,

	\begin{table}[t!]
		\centering
		\begin{tabular}{m{8cm}}
			\hline \hline
			\textbf{Algorithm 5} Parallel k-means++ Initialization \\
			1. Pick the initial data point uniformly at random from 1 to n. \\
			2. Initialize the n-length vector \texttt{Dist} where each element is the shortest distance between the data point $v_i$ and the current centroids. \\
			3. for $i$ = 2 to k \\
			\hspace{0.2cm} Compute the n-length vector $P$ such that  $P_j = \frac{\text{Dist}_j^2}{\sum_{l = 1}^n \text{Dist}_l^2}$  \\
			\hspace{0.2cm} Choose the $i^{\text{th}}$ centroid as the data point $x$ with probability $P_x$ \\
			\hspace{0.2cm} Compute the vector \texttt{newDist} such that  each $i^{\text{th}}$ element as the distance between the data point $v_i$ and the new centroid \\
			\hspace{0.2cm} Update \text{Dist} $\text{Dist}_j =  \text{minimum}(\text{Dist}_j, \text{newDist}_j )$\\
			\hline \hline	
		\end{tabular}
	\end{table}

	\par Step 3 in Algorithm 4 is the main loop that iteratively updates the labels of the data points and the corresponding centers of the clusters until convergence (or the maximum number of iterations is reached). Given the data points $V \in \mathbb{R}^{n \times d}$ and centroids $C \in \mathbb{R}^{k \times d}$, the pair-wise distance matrix $S \in \mathbb{R}^{n \times k}$ is computed as follows.
	\begin{equation}
	S_{ij} = \sum_{l = 1}^{d}(V_{il} - C_{jl})^2
	\end{equation}
	
	After expanding the right hand side, the distance matrix $S$ can be expressed as 
	\begin{equation}
	\label{reduced}
	S_{ij} = \sum_{l = 1}^{d}(V_{il})^2 + \sum_{l = 1}^{d}(C_{jl})^2 - 2\sum_{l = 1}^{d}V_{il}C_{jl}
	\end{equation}
	
	Hence, we compute two additional vectors $V_{norm} \in  \mathbb{R}^{n \times 1}$ and $C_{norm} \in  \mathbb{R}^{n \times 1}$, 
	\begin{equation}
	V_{norm}(i) = \sum_{l = 1}^{d}(V_{il})^2, 
	\end{equation}
	\begin{equation}
	C_{norm}(j) = \sum_{l = 1}^{d}(C_{jl})^2
	\end{equation}
	
	The matrix $S$ can be initialized as the sum of the corresponding elements in $V_{norm}$ and $C_{norm}$
	\begin{equation}
	S_{ij} = V_{norm}(i) + C_{norm}(j)
	\end{equation}
	
	The pair-wise distance matrix $S$ is then computed by level-3 BLAS function provided in the cuBLAS library.
	
	\begin{equation}
	S = S - 2VC^T
	\end{equation}
	
	\par For each data point, the new label is updated by as the index of centroid which has the minimum distance to the data point. Meanwhile, a global variable is maintained to record the number of label changes during the update.
	
	\par The new centroids are updated as the mean value of all the data points sharing the same label. To identify the points in each cluster, we sort the data points according to their new labels. Each GPU thread will then independently work on a consecutive portion of the sorted data points where most of these points share the same label. 
	
	\par
\begin{figure}[!t]
		\centering
		\includegraphics[width = .8\hsize]{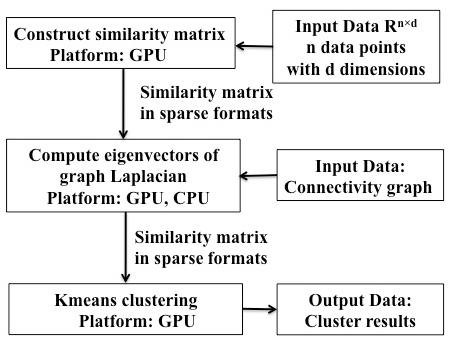}
		\caption{Parallel Implementation of Spectral Clustering}
		\label{wf}
	\end{figure}

 The entire workflow of our implementation is  summarized in Figure \ref{wf}. 
	
	\section{Evaluation}
	\subsection{Datasets}
	We evaluate our parallel implementation on several real-world and synthetic datasets. The  \textbf{ Diffusion Tensor Imaging (DTI)} dataset is given as a set of data points, each of which is characterized by a 90-dimensional array. The other datasets are specified by an undirected graph data where the edges are given by an edge list. The problem sizes and the numbers of clusters generated are shown in Table \ref{datasets}. A brief description of each dataset is given next. 
	
	\begin{itemize}
		\item \textbf{DTI}: The Diffusion Tensor Imaging(\textbf{DTI}) dataset is the brain image data of a subject chosen from a publicly accessible medical dataset provided by Nathan Kline Institute (NKI). The dataset captures the diffusion of the water 
		molecules in the brain tissues, which can be used to deduce information about the fiber connectivity in the human brain. After preprocessing steps \cite{jin2015}, the input data consists of $142K$ data points, each of which represents a 2mm$\times$2mm$\times$2mm brain voxel. The entire data points constitute the brain volume. Each data point is characterized by a 90-dimensional array representing the connectivity strength of the voxel to 90 brain regions (representing a segmentation of 
		the grey matter). The task is to cluster the voxels that share similar connectivity profiles. To facilitate the construction of the  similarity matrix, an edge list is provided which contains all pair of voxels that are within 4 millimeter distance.
		
		\item \textbf{FB}: This dataset is a dataset collected by a Facebook application. It contains the graph where each node represents an anonymous user and edges exist between users that share similar political interests\cite{snapnets}.
		
		\item \textbf{DBLP}: This dataset consists of a comprehensive co-authorship network in a computer science bibliography. The nodes represent the authors. Authors are connected if they coauthored at least one publication\cite{snapnets}. The dataset contains more than 5000 communities. Here we set the number of clusters to 500 for experimental purposes.
		
		\item \textbf{Syn200}: The synthetic dataset is randomly generated by the \textbf{stochastic block model} \cite{karrer2011stochastic}. The stochastic block model assumes that the data points are partitioned into $r$ disjoint subsets, $C_1, C_2, ..., C_r$. A symmetric $r \times r$ matrix $P$ is provided to model the inter-community edge probability. The synthetic sparse graph is randomly generated such that two nodes are connected with probability $p = 0.3$ if they are within the same cluster and $q = 0.01$ if they are in different clusters. 
	\end{itemize}
	
	\begin{table}[t!]
		\caption{Datasets}
		\label{datasets}
		\centering
		\begin{tabular}{|c|c|c|c|c|}
			\hline
			Dataset & Nodes & Edges & Clusters\\
			\hline
			DTI & 142541 & 3992290 & 500 \\ 
			\hline 
			FB & 4039 & 88234 & 10 \\
			\hline 
			DBLP & 317080 & 1049866 & 500 \\
			\hline 
			Syn200 & 20000 & 773388 & 200 \\
			\hline
		\end{tabular}
	\end{table}
	\subsection{Environment and Software}
	The computing environment is a heterogeneous CPU-GPU platform with CPU and GPU specifics shown in Table \ref{Specifics}. The software and packages used are as follows,
	
	\begin{itemize}
		\item \textbf{Matlab}: Matlab is a high-level language that provides interactive programing environment,  which is widely used by  scientists and engineers. The version of Matlab used for our implementation is 2015a. The sparse matrix representation and operations are the built-in functions. The k-means clustering is the function in Statistical and Machine Learning toolbox.
		
		\item \textbf{Python}: Python software packages, such as Numpy, Scipy and sklearn, are  popular tools to perform scientific computations.
		The version of Python binary for our implementation is 2.7.11. The sparse representation and functions to solve the eigenvalue problems are from \emph{Scipy} package. The k-means clustering function is from \emph{sklearn.cluster} module. The module versions are Numpy-1.10.4, Scipy-0.16.1 and sklearn-0.17 respectively.
	\end{itemize}
	
	\par  Linear algebra and numeric functions are by default multi-threaded in Matlab on multicore and multiprocessor machines \footnote{http://www.mathworks.com/discovery/matlab-multicore.html}. In addition, the Python packages are built on highly optimized CPU-based BLAS routines, some of which have been accelerated using multi-threaded programming. 
	
	\subsection{Performance Analysis} 
	We measure the running time of our  spectral clustering algorithm on the three components separately: 1) computation of the similarity matrix; 2) sparse matrix eigensolver; and 3) the k-means clustering algorithm. For the CUDA implementation, we measure the time costs that include both the computational time as well as the extra time for library initialization time and data communication. Specifically, we evaluate the performance of each of the following components:
	\begin{itemize}
		\item \textbf{Computation of the  similarity matrix}:
		\begin{itemize}
			\item initialize CUDA libraries.
			\item transfer data and edge list from CPU to GPU. 
			\item construct the similarity matrix.
		\end{itemize}
		\item \textbf{Sparse matrix eigensolver}: 
		\begin{itemize}
			\item data communication between CPU and GPU;
			\item computation of the  eigenvectors;
			\item transfer of the eigenvectors from CPU to GPU.
		\end{itemize} 
		\item \textbf{K-means clustering}:
		\begin{itemize}
			\item perform the k-means clustering;
			\item tranfer the clustering result from GPU to CPU.
		\end{itemize}
	\end{itemize}
	
	\par
\begin{figure}[!t]
		\centering
		\includegraphics[width = .8\hsize]{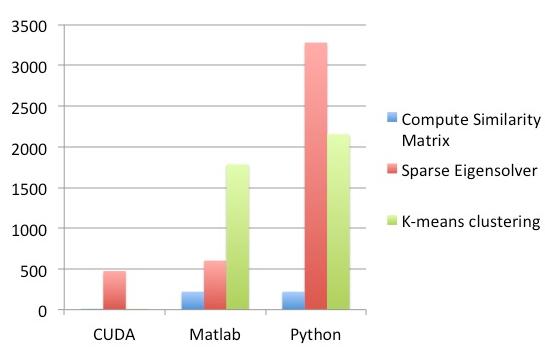}
		\caption{Time Costs of Spectral Clustering on DTI Dataset}
		\label{DTIfigure}
	\end{figure}

 Figure \ref{DTIfigure}. and Table \ref{DTItable}. show the time costs of each step corresponding to the \textbf{DTI} dataset. 
	
	\begin{table}[t!]
		\caption{Running Time of Spectral Clustering on DTI Dataset}
		\label{DTItable}
		\centering
		\begin{tabular}{|c|c|c|c|}
			\hline
			Time/s & CUDA & Matlab & Python\\
			\hline 
			Compute Similarity Matrix & \textbf{0.0331} & 221.249 & 220.880 \\
			\hline 
			Sparse Eigensolver & \textbf{475.442} & 603.165 & 3281.973 \\
			\hline 
			K-means Clustering & \textbf{5.407}& 1785.17 & 2154.7818\\
			\hline 
		\end{tabular}
	\end{table}
	
	\par It is clear that our CUDA implementation significantly outperforms the currently fastest known Matlab and Python implementations at each step. Since the computation of the similarity matrix is highly parallel, the CUDA implementation achieves linear speedups by taking advantage of the GPU with thousands of threads computing the cross correlation coefficients concurrently. For the Matlab and Python implementations, the results are based on the serial implementation which loops over the edge list and computes the correlation coefficient explicitly using the built-in function. We also tested an alternative implementation which takes advantage of \emph{vectorization} techniques that recast the loop-based operation into matrix and vector operations. The optimized Matlab and Python implementationd take $5.753s$ and $6.271s$ trespectively to compute the similarity matrix.  
	
	\par Both Matlab and Python packages utilize the \emph{reverse communication interfaces} of ARPACK to compute the eigenvectors of large-scale symmetric matrix, and hence all of the three implementations share similar procedures and interfaces. The  basic difference is related to  the function to compute the sparse matrix-vector multiplication. Our CUDA implementation utilizes the GPU and the cuSPARSE library to compute the multiplication while Matlab and Python utilize their built-in routines. Since the GPU performs significantly better than the CPU on BLAS operations \cite{cullinan2013computing}, the CUDA implementation achieves better performance than Matlab and Python even with the communication overhead. However, since the time complexity of implicitly restarted Lanczos method is approximately $O(m^3 + nm^2)$, the time spent on the \emph{reverse communication interfaces} scales relatively poorly, which may become the most computationally expensive part when $k$ is large. 
	
	\par As for the  \emph{kmeans clustering} algorithm, our CUDA implementation achieves more than 300x speedup over the Matlab and Python implementations. The running time of  this step depends on the centroid initialization. The CUDA and Python implementations utilize the k-means++ initialization, which leads to fewer number of iterations in general  than Matlab. Moreover, in the CUDA implementation, the process of transforming the computation of the pair-wise distance matrix to the BLAS operations significantly accelerates the running time of the algorithm.

	\par
\begin{figure}[!t]
		\centering
		\includegraphics[width = .8\hsize]{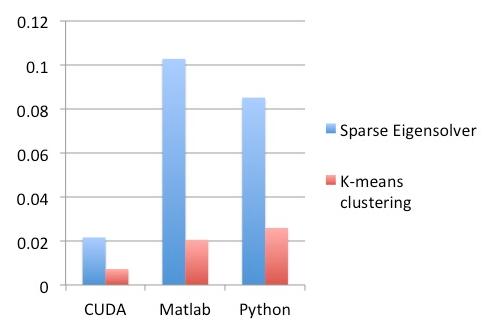}
		\caption{Time Costs of Spectral Clustering on FB Dataset}
		\label{FBfigure}
	\end{figure}

\begin{figure}[!t]
		\centering
		\includegraphics[width = .8\hsize]{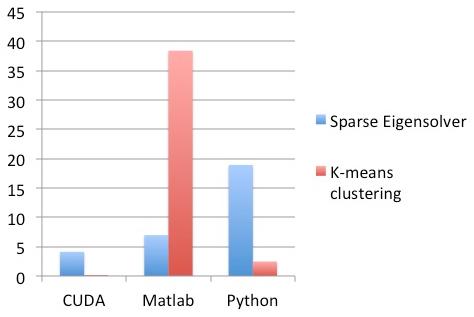}
		\caption{Time Costs of Spectral Clustering on Syn200 Dataset}
		\label{synfigure}
	\end{figure}

\begin{figure}[!t]
		\centering
		\includegraphics[width = .8\hsize]{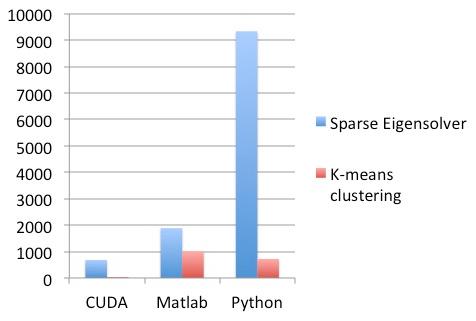}
		\caption{Time Costs of Spectral Clustering on dblp Dataset}
		\label{dblpfigure}
	\end{figure}

 The performance results for the graph datasets (\textbf{FB}, \textbf{Syn200}, \textbf{dblp}) are shown in Table \ref{FBtable} through Table \ref{dblptable} and Figure \ref{FBfigure} through Figure \ref{dblpfigure}. Similar to the previous results, our CUDA implementation achieves the best performance among  the three implementations at each step. However, the speedup ratio  depends on the specific problem size.
	
	\begin{table}[t!]
		\caption{Running Time of Spectral Clustering on FB Dataset}
		\label{FBtable}
		\centering
		\begin{tabular}{|c|c|c|c|}
			\hline
			Time/s & CUDA & Matlab & Python\\
			\hline 
			Sparse Eigensolver & \textbf{0.0216} & 0.1027 & 0.0851 \\
			\hline 
			K-means Clustering & \textbf{0.007251}& 0.0205 & 0.0259\\
			\hline 
		\end{tabular}
	\end{table}
	
	The FB dataset contains a very small graph with 4039 nodes and involves very few clusters $k = 10$. Because the number of clusters is small, the most expensive computation of \emph{sparse eigensolver} is the sparse matrix-vector multiplication. Therefore for this step,the  CUDA implementation achieves around 5x speedup over the other implementations. For the  \emph{k-means clustering} step, the CUDA implementation shows only a minor speedup by a factor of around 4x.
	
	\begin{table}[t!]
		\caption{Running Time of Spectral Clustering on Syn200 Dataset}
		\label{syntable}
		\centering
		\begin{tabular}{|c|c|c|c|}
			\hline
			Time/s & CUDA & Matlab & Python\\
			\hline 
			Sparse Eigensolver & \textbf{4.1153} & 6.9531 & 18.915\\
			\hline 
			K-means Clustering & \textbf{0.02478} & 38.3728 & 2.4719\\
			\hline 
		\end{tabular}
	\end{table}
	
	\par The Syn200 dataset contains a medium-sized synthetic graph with 200 clusters. The CUDA implementation achieves a slight improvement in computing the eigenvectors since the performance is mainly constrained by the CPU-based routines. For the  of k-means clustering step, the CUDA implementation achieves over 100x speedup.

	\begin{table}[t!]
		\caption{Running Time of Spectral Clustering on dblp Dataset}
		\label{dblptable}
		\centering
		\begin{tabular}{|c|c|c|c|}
			\hline
			Time/s & CUDA & Matlab & Python\\
			\hline 
			Sparse Eigensolver & \textbf{682.643} & 1885.2303 & 9338.31\\
			\hline 
			K-means Clustering & \textbf{1.79456} & 1012.92 & 719.686\\
			\hline 
		\end{tabular}
	\end{table}

	\par The dblp dataset contains a large-scale graph with 500 clusters. Both Matlab and Python implementations perform poorly for such a  problem size. Our CUDA implementation achieve around 3x speedup in \emph{sparse eigensolver} in spite of the fact that the performance is still constrained by the CPU-based interfaces. In the  k-means clustering step, the CUDA implementation achieves over 400x speedup.
	
	\par Table \ref{comp} shows a comparison between data communication time and computation time for the CUDA implementation on each of our four datasets. The data communication time includes 1) input data transfered from CPU to GPU; 2) data communication between CPU and GPU during the execution of the eigensolver stage; 3) output results that are transferred from GPU to CPU.  Given that the bandwidth remains constant during the execution of the algorithm, the time complexity of data communication is  $O(n^2 + m \times \# + nk)$ depending on the sparsity ratio of the similarity matrix and the number of Arnoldi iterations $\#$ n; the time complexity of computation is $O(nd^2 + O(nm^2) \times \# + O(n^2k))$. Therefore we expect the data communication time to be less than the computational time as in fact illustrated in the Table \ref{comp}, especially for large-scale problems.
	
	\begin{table}[t!]
		\caption{Comparison  Between Data Communication Time and Computation Time}
		\label{comp}
		\centering
		\begin{tabular}{|c|c|c|}
			\hline
			Time/s &  Communication & Computation \\
			\hline 
			DTI & 2.248 & 475.213 \\
			\hline 
			FB  & 0.002131 & 0.02635\\
			\hline 
			DBLP & 2.731 & 680.31 \\
			\hline 
			Syn200 & 0.0741 & 3.8201\\
			\hline 
		\end{tabular}
	\end{table}

	\par In conclusion, our CUDA implementation always achieves better performance than Matlab and Python implementations for each step. The speedup ratio largely depends on the specific problem size. Our traget applications involve  problems with a large number of clusters. Our implementation achieves significant speedups for the steps of \emph{computing the similarity matrix} 
	and the \emph{k-means clustering} due to the massive computational power of GPU. Moreover, we always achieve some speedups for the \emph{sparse eigensolver} step by accelerating the  computations involving matrix-vector multiplications.
	
	\section{Conclusion}
	We presented a high performance implementation of the spectral clustering algorithm on CPU-GPU platforms. Our implementation leverages the GPU to accelerate highly parallel computations and Basic Linear Algebra Subprograms (BLAS) operations. We focused  on the acceleration of the three major steps of the spectral clustering algorithm: 1) construction of the similarity matrix; 2) computation of eigenvectors for large-scale similarity matrices; 3) k-means clustering algorithm. We believe that we are the first to accelerate the large-scale eigenvector computation by combining the interfaces of traditional CPU-based software packages ARPACK and GPU-based CUDA library. Such a combination  achieves good speedups compared to other CPU-based software. We deploy a smart seeding strategy and utilize BLAS operations to implement the fast k-means clustering algorithm. Our implementation is shown to achieve significant speedup compared to Matlab and Python software packages, especially for large-scale problems.
	
	\section*{Acknowledgment}
	We gratefully acknowledge funding provided by The University of Maryland/Mpowering the State through the Center for Health-related Informatics and Bioimaging (CHIB) and the NVIDIA Research Excellence
	Center at the University of Maryland.
	
	\printbibliography

\end{document}